\documentclass[seceq]{ptptex}
\usepackage{graphicx}
\usepackage{wrapft}
\usepackage{bm}
\def\Vec#1{\mbox{\boldmath $#1$}}

\def\itmb{\begin{itemize}}
\def\itme{\end{itemize}}
\def\enmb{\begin{enumerate}}
\def\enme{\end{enumerate}}
\def\eqnb{\begin{equation}}
\def\eqne{\end{equation}}
 
\def\PTP{Prog. Theor. Phys.(Kyoto)}

\def\NPB{{Nucl. Phys.} { B}}
\def\PLB{{Phys. Lett.} B}
\def\PRL{Phys. Rev. Lett.}
\def\PRD{{Phys. Rev.} D}

\title{On the roles of triality in the infrared QCD}

\author{
Sadataka \textsc{Furui}%
}

\inst{
School of Science and Engineering, Teikyo University\\
 Utsunomiya 320-8551, Japan
}

\abst{%
In the QCD analysis, when quarks are expressed in quaternion basis, the product of quaternions are expressed by octonions and the octonion posesses the triality symmetry. Roles of triality in the gluon self energy calculation and finite temperature QCD is explained. 
}

\begin{document}
\newcommand{\ttbs}{\char'134}
\newcommand{\Slash}[1]{\ooalign{\hfil/\hfil\crcr$#1$}}
\maketitle
\section{Introduction}
\label{sect1}
In \cite{SF10}, I discussed a possible solution of the discrepancy of the critical flavor number ${N_f}^c$ for a presence of infra red (IR) fixed point of QCD in the Schr\"odinger functional lattice simulation, which implies ${N_f}^c\geq 10$ and a hint of the presence of the fixed point in $N_f=2+1$ system in the polarized electron-proton scattering data of JLab \cite{DBCK06,DBCK08}.
In the extraction of effective coupling $\alpha_{g1}(Q^2)$ of QCD from electron scattering data, Bjorken sum rule 
\begin{equation}
\int_0^1 dx[g_1^{ep}(x,Q^2)-g_1^{en}(x,Q^2)]=\frac{1}{6}\left|\frac{g_A}{g_V}\right|[1-\frac{\alpha_{g1}(Q^2/\Lambda^2_{g1})}{\pi}]
\end{equation}
was adopted.

 The experiment shows that near IR limit $\alpha_{g1}(Q^2)\sim \pi$, but there is an argument that the JLab experiment does not imply proximity of the conformal window at real QCD and deny discrepancy. An aim of this paper is to explain why I think there are discrepancies, and large $N_f$ appearance of conformal symmetry in analytical and in lattice Shr\"odinger functional simulation is relevant to the experimental data.

Presence of infrared fixed point and proximity of the conformal window was proposed by Banks and Zaks \cite{BZ82}. In order to prove the presence of infrared fixed point, a power low behavior of the correlator that satisfy Callan-Symanzik equation is a sufficient condition, but not a necessary condition.  In the theory of Grunberg on the effective charge \cite{Gr89}, renormalization scheme for the effective coupling is not that of Callan-Symanzik approach, but that of Gell-Mann-Low approach.  An effective charge $\alpha_s^f(r)$ defined in the work is associated with
\begin{equation}
F(r)=\frac{dV}{dr}=\frac{\alpha_s^f(r)}{r^2},
\end{equation}
instead of an effective charge $\alpha_s^p(r)$ associated with 
\begin{equation}
V(r)=-\frac{\alpha_s^p(r)}{r}.
\end{equation}

In contrast to $\alpha_s^f(r)$,
a single-valued $\beta$ function description does not work for $\alpha_s^p(r)$
over the whole range of $r$, since $V(r)$ and hence $\alpha_s^p(r)$ change sign  somewhere between Coulomb and confinement regions. The Fourier transform of renormalization invariant coupling $\displaystyle r\to \frac{1}{Q}$ is not always applicable. However, Shirkov \cite{Sh03} conjectured that IR asymptotics of $\alpha_{SF}(Q^2)$ of Schr\"odinger functional approach of ALPHA collaboration\cite{ALPHA95} would be finite or slightly singular($\displaystyle \sim\frac{1}{Q}$).  
In the analytic perturbation theory \cite{GGK98}, in which a single-valued $\beta$ function that avoids Landau singularity is used in the 3-loop calculation, presence of infrared fixed point for $N_f>10$, independent of renormalization scheme  was shown.  

Phenomenologically, walking behavior of the QCD effective coupling is a sign of the presence of infrared fixed point. In Dyson-Schwinger approach, with constant coupling constant in the infrared region, a model that is consistent with asymptotic freedom and chiral symmetry breaking was constructed \cite{Hi84} and nearly constant QCD coupling in infrared region is also shown in more sophisticated Dyson-Schwinger approach \cite{FA03,AS01}.

In \cite{SF10}, I discussed importance of self-dual gauge field or instantons in IR QCD. D'Adda and Di Vecchia \cite{DD78} showed that in the supersymmetric Yang-Mills theory, cancellation of the fluctuation from the fermion sector and boson sector occurs, and that the one loop correction around any self-dual instantons is just given by the zero modes of the gluon, fermion and ghost.
Proximity of the conformal window and the behavior of the effective coupling that approaches a constant in the infrared in supersymmetric QCD was discussed in \cite{GG99}.   

Lattice simulation in Schr\"odinger functional method also suggests that critical number of fermions $N_f^c$ for opening the conformal window is $\sim 10$\cite{App08}. When one accepts the argument of Banks and Zaks \cite{BZ82} and the hint of presence of infrared fixed point in the lattice results of the effective coupling in Coulomb gauge in 2+1 flavors \cite{SF10} and in the Dyson-Schwinger results of the effective coupling in Landau gauge \cite{FA03}, one can conclude that there is discrepancy.  In \cite{SF10}, I discussed that the discrepancy could be resolved when one analyze fermions in quaternion basis. I explain in this paper the role of triality in the IRQCD.
 
Contents of this paper is as follows. I explain the triality transformation in the gluon self energy in sect.2,  and show how the triality plays a role in finite temperature QCD in sect. 3.  A discussion is given in sect.4.

\section{Triality of the gluon self energy}
Quaternions are generalization of complex number ${\mathcal C}={\mathcal R}+i{\mathcal R}$, which are expressed as $q=w+x{\Vec i}+y{\Vec j}+z{\Vec k}$. Automorphism group of ${\mathcal H}={\mathcal R}+{\mathcal R}^3$ is SO(3).

A new imaginary unit $l$ that anticommutes with the bases of quaternions ${\Vec i},{\Vec j},{\Vec k}$ compose octonions ${\mathcal O}={\mathcal H}+l{\mathcal H}$. Automorphism group of ${\mathcal O}={\mathcal R}+{\mathcal R}^7$ is not SO(7), but exceptional Lie group $G_2$. It contains tensor product of three ${\mathcal R}^7$ bases and three vectors.
The triality automorphism is a transformation that rotates 24 dimensional bases defined by Cartan\cite{Cartan66}.

\[
\{\xi_0, \xi_1, \xi_2, \xi_3, \xi_4\},\quad
\{\xi_{12}, \xi_{31}, \xi_{23}, \xi_{14}, \xi_{24}, \xi_{34}\},\quad
\{\xi_{123}, \xi_{124}, \xi_{314}, \xi_{234}, \xi_{1234}\}
\]
\[
\{x^1, x^2, x^3, x^4\}, \quad \{x^{1'}, x^{2'}, x^{3'}, x^{4'}\}
\]

The trilinear form in these bases is
\begin{eqnarray}
{\mathcal F}&=&\phi^T CX\psi=x^1(\xi_{12}\xi_{314}-\xi_{31}\xi_{124}-\xi_{14}\xi_{123}+\xi_{1234}\xi_1)\nonumber\\
&+&x^2(\xi_{23}\xi_{124}-\xi_{12}\xi_{234}-\xi_{24}\xi_{123}+\xi_{1234}\xi_2)\nonumber\\
&+&x^3(\xi_{31}\xi_{234}-\xi_{23}\xi_{314}-\xi_{34}\xi_{123}+\xi_{1234}\xi_3)\nonumber\\
&+&x^4(-\xi_{14}\xi_{234}-\xi_{24}\xi_{314}-\xi_{34}\xi_{124}+\xi_{1234}\xi_4)\nonumber\\
&+&x^{1'}(-\xi_{0}\xi_{234}+\xi_{23}\xi_{4}-\xi_{24}\xi_{3}+\xi_{34}\xi_2)\nonumber\\
&+&x^{2'}(-\xi_{0}\xi_{314}+\xi_{31}\xi_{4}-\xi_{34}\xi_{1}+\xi_{14}\xi_3)\nonumber\\
&+&x^{3'}(-\xi_{0}\xi_{124}+\xi_{12}\xi_{4}-\xi_{14}\xi_{2}+\xi_{24}\xi_1)\nonumber\\
&+&x^{4'}(\xi_{0}\xi_{123}-\xi_{23}\xi_{1}-\xi_{31}\xi_{2}-\xi_{12}\xi_3)
\end{eqnarray}
There are three semi-spinors which have a quadratic form which is invariant with respect to the group of rotation
\begin{eqnarray}
\Phi={^t\phi} C\phi&=&\xi_0\xi_{1234}-\xi_{23}\xi_{14}-\xi_{31}\xi_{24}-\xi_{12}\xi_{34}\\
\Psi={^t\psi} C\psi&=&-\xi_1\xi_{234}-\xi_{2}\xi_{314}-\xi_{3}\xi_{124}+\xi_{4}\xi_{123}\end{eqnarray}
and the vector
\begin{equation}
F=x^1 x^{1'}+x^2 x^{2'}+x^3 x^{3'}+x^4 x^{4'}
\end{equation}

With use of quaternion bases $1,{\Vec i},{\Vec j}, {\Vec k}$, the spinors $\phi$ and $C\phi=\phi'$ are defined as
\begin{eqnarray}
\phi&=&\xi_0+\xi_{14}{\Vec i}+\xi_{24}{\Vec j}+\xi_{34}{\Vec k}\\
C\phi&=&\xi_{1234}-\xi_{23}{\Vec i}-\xi_{31}{\Vec j}-\xi_{12}{\Vec k}.
\end{eqnarray}
Similarly, $\psi$ and $C\psi=\psi'$ are defined as
\begin{eqnarray}
\psi&=&\xi_4+\xi_1{\Vec i}+\xi_2{\Vec j}+\xi_3{\Vec k}\\
C\psi&=&\xi_{123}-\xi_{234}{\Vec i}-\xi_{314}{\Vec j}-\xi_{124}{\Vec k}.
\end{eqnarray}

When the trilinear form is given, one can construct 3-loop self energy diagram of the vector field $x_1, x_2, x_3$ and $x_4$ as shown in Figs.1-17.

\begin{figure}
\begin{minipage}[b]{0.47\linewidth}
\begin{center}
\includegraphics[width=6cm,angle=0,clip]{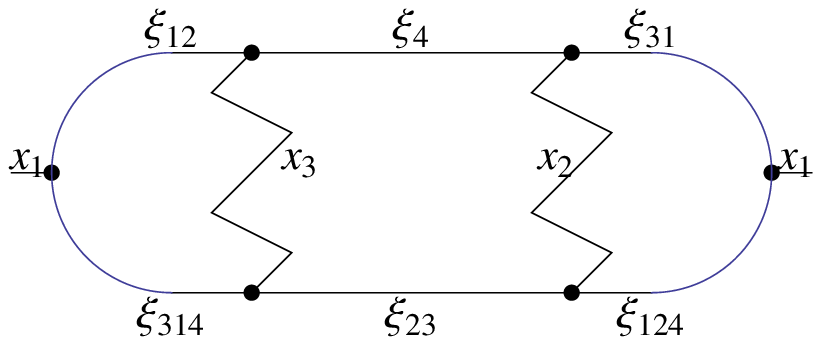}
\caption{g611a}
\label{11a}
\end{center}
\end{minipage}
\hfill
\begin{minipage}[b]{0.47\linewidth}
\begin{center}
\includegraphics[width=6cm,angle=0,clip]{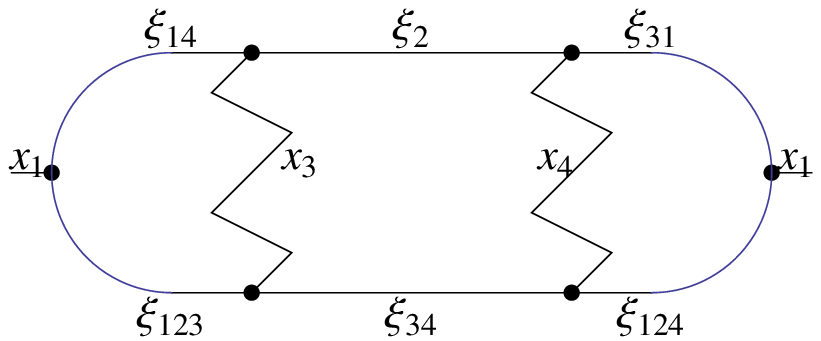}
\caption{g611b}
\label{11b}
\end{center}
\end{minipage}
\end{figure}
\begin{figure}
\begin{minipage}[b]{0.47\linewidth}
\begin{center}
\includegraphics[width=6cm,angle=0,clip]{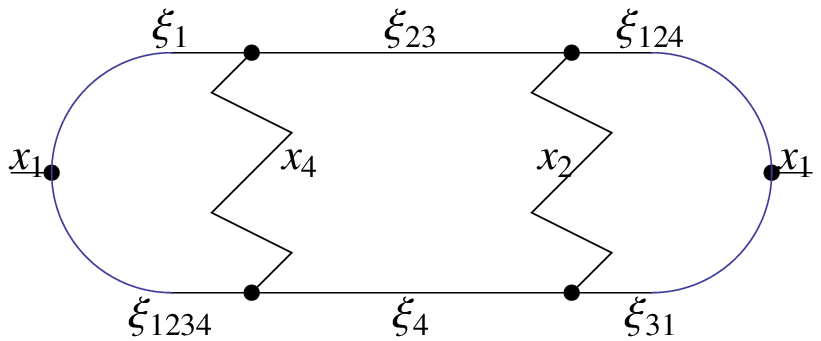}
\caption{g611c}
\label{11c}
\end{center}
\end{minipage}
\hfill
\begin{minipage}[b]{0.47\linewidth}
\begin{center}
\includegraphics[width=6cm,angle=0,clip]{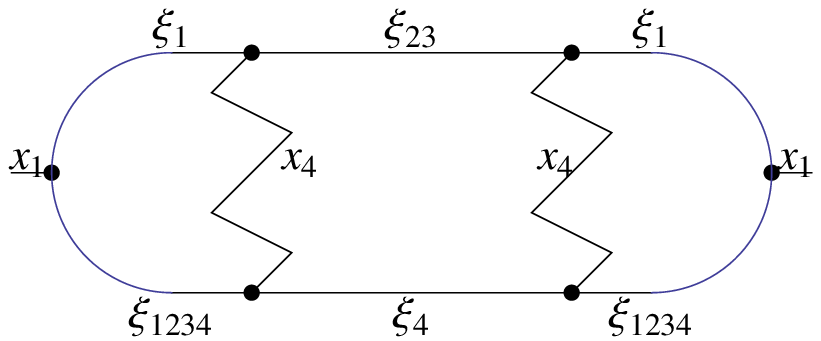}
\caption{g611d}
\label{11d}
\end{center}
\end{minipage}
\end{figure}

\begin{figure}
\begin{minipage}[b]{0.47\linewidth}
\begin{center}
\includegraphics[width=6cm,angle=0,clip]{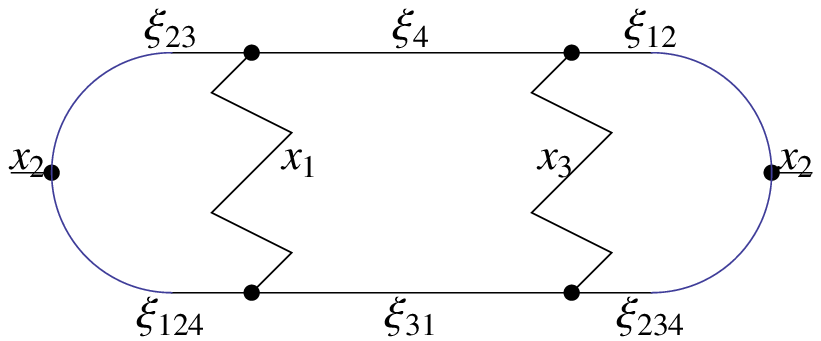}
\caption{g622a}
\label{22a}
\end{center}
\end{minipage}
\hfill
\begin{minipage}[b]{0.47\linewidth}
\begin{center}
\includegraphics[width=6cm,angle=0,clip]{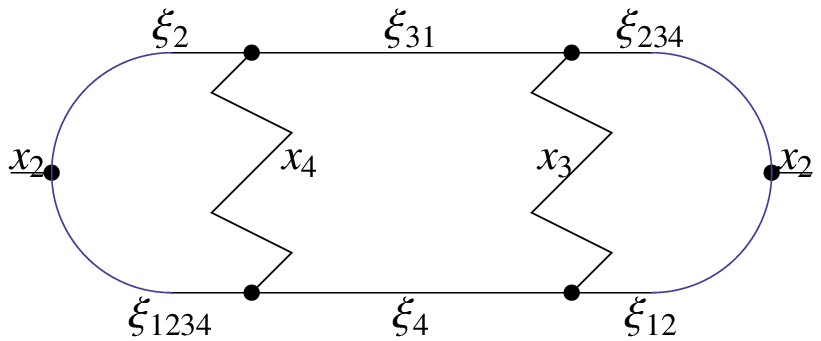}
\caption{g622b}
\label{22b}
\end{center}
\end{minipage}
\end{figure}
\begin{figure}
\begin{minipage}[b]{0.47\linewidth}
\begin{center}
\includegraphics[width=6cm,angle=0,clip]{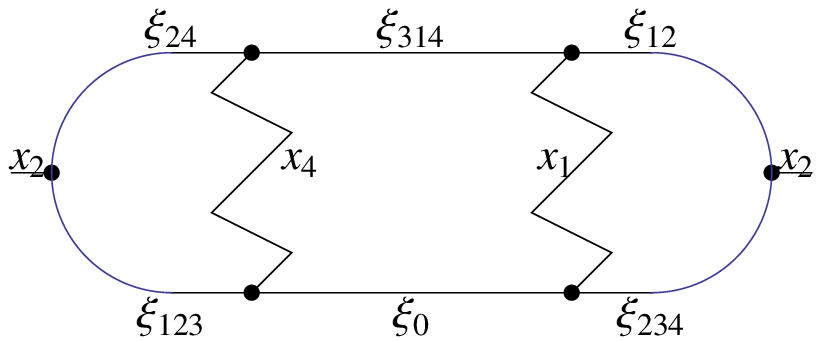}
\caption{g622c}
\label{22c}
\end{center}
\end{minipage}
\hfill
\begin{minipage}[b]{0.47\linewidth}
\begin{center}
\includegraphics[width=6cm,angle=0,clip]{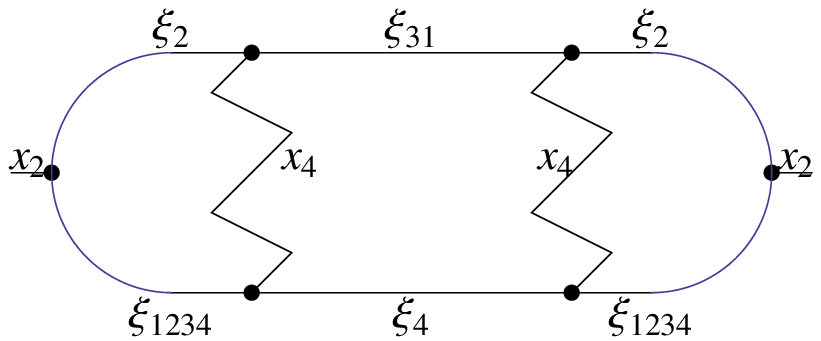}
\caption{g622d}
\label{22d}
\end{center}
\end{minipage}
\end{figure}

\begin{figure}
\begin{minipage}[b]{0.47\linewidth}
\begin{center}
\includegraphics[width=6cm,angle=0,clip]{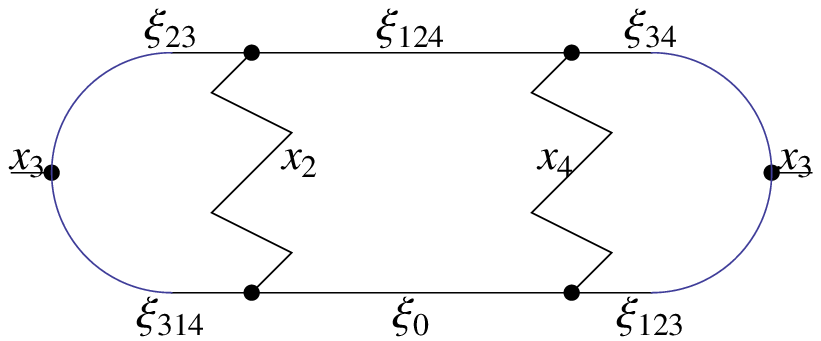}
\caption{g633a}
\label{33a}
\end{center}
\end{minipage}
\hfill
\begin{minipage}[b]{0.47\linewidth}
\begin{center}
\includegraphics[width=6cm,angle=0,clip]{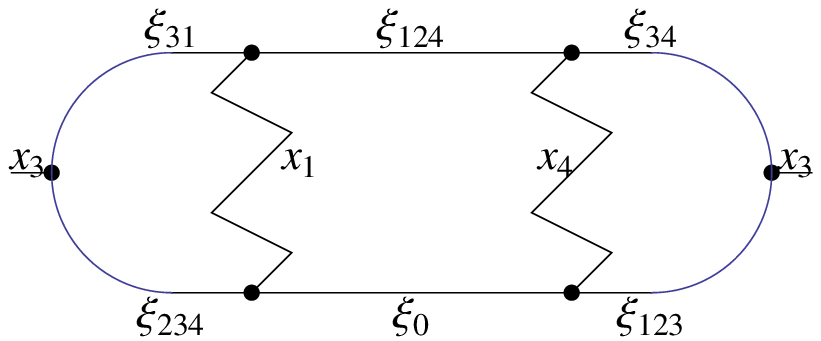}
\caption{g633b}
\label{33b}
\end{center}
\end{minipage}
\end{figure}
\begin{figure}
\begin{minipage}[b]{0.47\linewidth}
\begin{center}
\includegraphics[width=6cm,angle=0,clip]{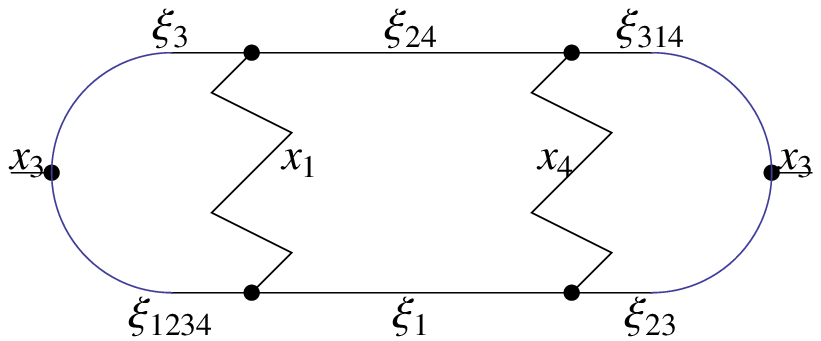}
\caption{g633c}
\label{33c}
\end{center}
\end{minipage}
\hfill
\begin{minipage}[b]{0.47\linewidth}
\begin{center}
\includegraphics[width=6cm,angle=0,clip]{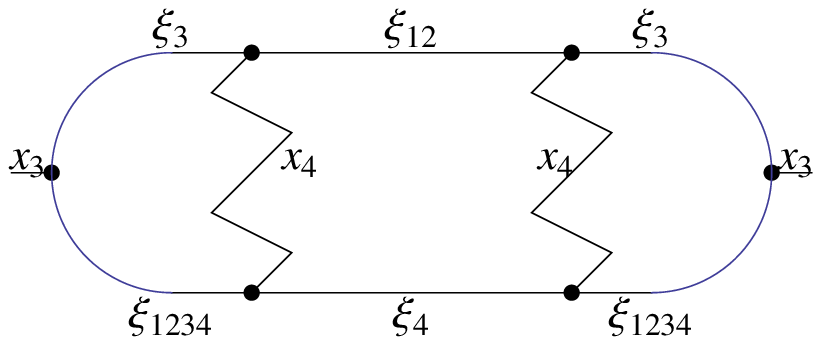}
\caption{g633d}
\label{33d}
\end{center}
\end{minipage}
\end{figure}

\begin{figure}
\begin{minipage}[b]{0.47\linewidth}
\begin{center}
\includegraphics[width=6cm,angle=0,clip]{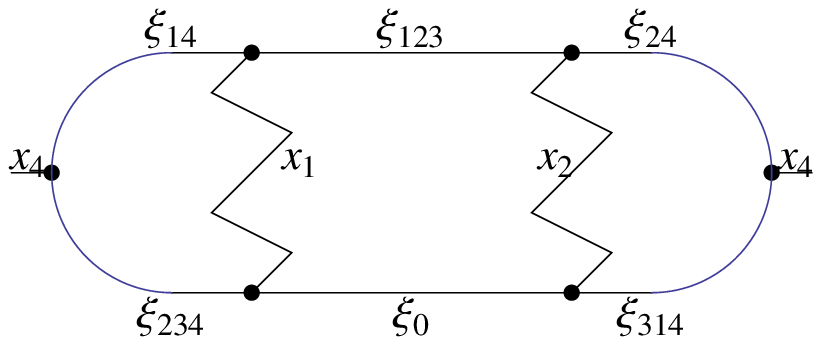}
\caption{g644a}
\label{44a}
\end{center}
\end{minipage}
\hfill
\begin{minipage}[b]{0.47\linewidth}
\begin{center}
\includegraphics[width=6cm,angle=0,clip]{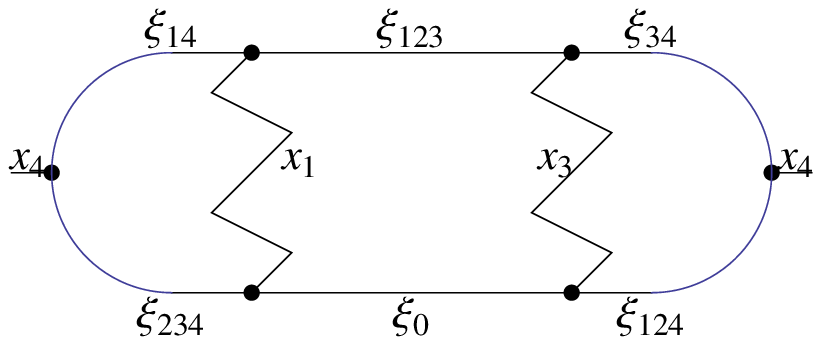}
\caption{g644b}
\label{44b}
\end{center}
\end{minipage}
\end{figure}
\begin{figure}
\begin{minipage}[b]{0.47\linewidth}
\begin{center}
\includegraphics[width=6cm,angle=0,clip]{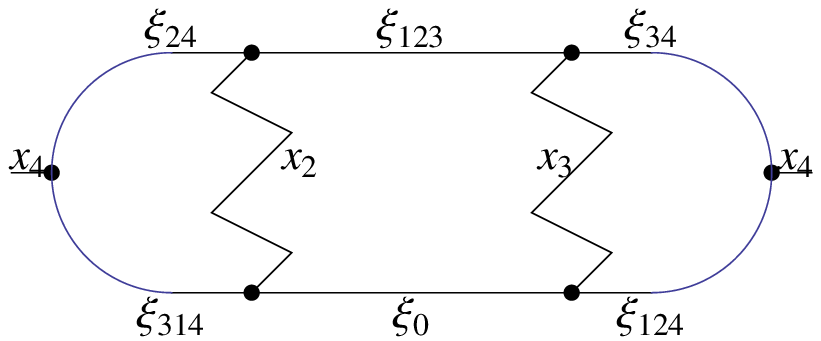}
\caption{g644c}
\label{44c}
\end{center}
\end{minipage}
\hfill
\begin{minipage}[b]{0.47\linewidth}
\begin{center}
\includegraphics[width=6cm,angle=0,clip]{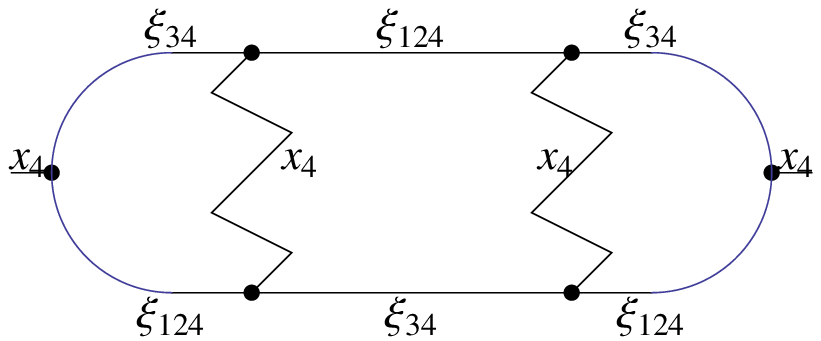}
\caption{g644d}
\label{44d}
\end{center}
\end{minipage}
\end{figure}

\begin{figure}
\begin{center}
\includegraphics[width=6cm,angle=0,clip]{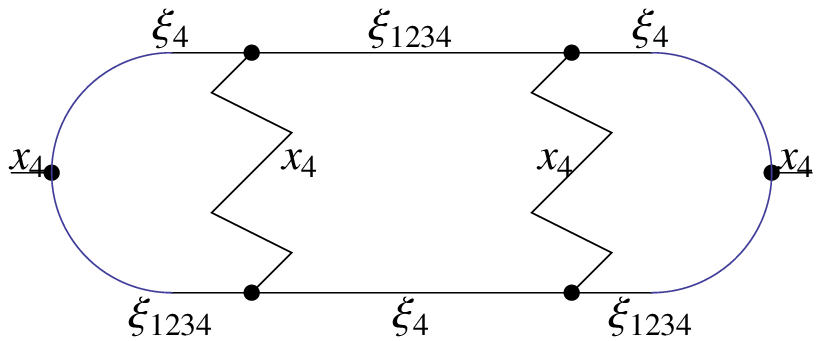}
\caption{g644e}
\label{44e}
\end{center}
\end{figure}

In these diagrams, $x_i$ is a self-dual gauge field that stands for $x^i$ and its conjugate $x^{i'}$. At each quark-gluon vertices, a factor +1 or -1 is multiplied, but  the number of vertices where -1 is multiplied in each diagram is even in all the 17 diagrams, and so there is no problem of relative sign among the diagrams.
In the figures, the diagram whose left-hand side and the right hand side are interchanged is not shown. Hence, there are 7 diagrams of equal signs for a longitudinal direction of the gluon along $x,y$ and $z$ axis, respectively, and 8 diagrams for longitudinal direction of the gluon along the $t$ axis.

Since electron number and muon number are conserved in experiments, I expect the quark weak-boson coupling preserves triality.  But if the quark-gluon coupling is triality blind, the effetive number of quark flavors contributes in the gluon self energy via quark loops, the self energy becomes three times larger. In the 17 diagrams, final gluon could couple to quark-anti quark pairs belonging to a different triality sector than that of the initial gluon.
 The difference of quark weak-boson coupling and quark-gluon coupling is expected for solving the fine-tuning problem in the grand unified theory \cite{Ran07}.

\section{Triality in finite temperature QCD}
In finite temperature QCD, there is the magnetic mass problem \cite{IKRV06}, i.e. in the limit of $P\to 0$, the argument of $\log(1+\Pi_T(P)/P^2)$ becomes negative, resulting in an unwanted imaginary contribution in the QCD pressure. For $p_0=0$, the pole is determined from \cite{KK81}  
\begin{equation}
p^2+\Pi_T(p_0=0,p)=p^2-g^2N_c T\frac{8+(\xi+1)^2}{64}p \label{pressure}
\end{equation}
where $\xi$ is the gauge parameter of covariant gauges, $N_c$ is the number of colors. 
At $p_0=0$ and in the limit of $\bf p\to 0$, the 3-loop diagrams Fig.1-17 expected to yield the self energy $g^6T^3/m$ \cite{Li80}, where $m$ is the infrared cut off. If one takes the electric mass $m=m_{el}\sim gT$ and the magnetic mass $m_{mag}\sim g^2 T$, a sum of the loops gives $m_{mag}^2\sim cg^4 T^2$ and $m_{mag}^3 T\sim g^6 T^4$. 
These contributions are expected to play roles in cancelling the unwanted pole at $p=g^2N_c T\frac{8+(\xi+1)^2}{64}$ from the inverse of eq.(\ref{pressure}).

 In a perturbative analysis of finite temperature QCD, the inverse gluon propagator goes like \cite{Li80}
\begin{equation}
p^2+a_1 g^2 T p+a_2 g^4 T^2+a_3\frac{g^6 T^3}{p}+\cdots.
\end{equation}
However, there is no systematic way to evaluate $a_3$. Whether the 3-loop diagram with exchange of two self-dual gluon fields dominates in $g^6$ term, as expected from the conjecture of \cite{DD78} is to be investigated.

\section{Discussion}
In this section I discuss on the technicolor theory \cite{CL84} and on unparticle physics \cite{Ge07}.

The technicolor was introduced from an estimation of the gluon self energy diagram as Fig.1-17 but with quark loops in the standard model, which yields in Minkowski space,
\begin{equation}
\Pi^{ij}_{\mu\nu}(p)=\delta^{ij}\left(\frac{g}{2}\right)^2(g_{\mu\nu}-p_\mu p_\nu/p^2)p^2\Pi(p^2).
\end{equation}
Summing up the bubbles the propagator becomes
\begin{equation}
\delta^{ij}\frac{g_{\mu\nu}-p_\mu p_\nu/p^2}{p^2[1-g^2\Pi(p^2)/4]}
\end{equation}
Assuming that the chiral symmetry is realized in the Goldstone mode, one expects $\Pi(p^2)=f_\pi^2/p^2$, where $f_\pi$ is the pion decay constant. A comparison
with experiment shows that $M^2=g^2f_\pi^2/4$ is too small and the technicolor interaction $M_W^2=g_{TC}^2F_\pi^2/4$ with scale parameter $\Lambda_{TC}\sim 1$TeV was introduced. When a gluon is triality blind, the bubble summation of the gluon propagator would change, and the technicolor theory would be modified.

Unparticle physics formulated by Georgi \cite{Ge07} starts from Lagrangian, which contains a nontrivial IR fixed point, and an assumption of presence of unparticle $\mathcal U$ of scale dimension $d_{\mathcal U}$, which appear in
\begin{equation}
q+\bar q\to gluon +{\mathcal U},  \qquad q+ gluon\to q+{\mathcal U}.
\end{equation} 
If the unparticle can be interpreted as a quark-anti quark pair in a different triality sector as the original, the detection of the unparticle by electro-magnetic detector would be difficult.

 To establish the IRQCD in quaternion basis and confirm importance of the triality,  both lattice simulation and analytical calculation are necessary. Extension of our analysis of domain wall quark lattice simulation of $16^3\times 32\times 16$ to larger lattices and an estimation of the behavior in the continuum limit is under way.

\end{document}